\documentclass[aps,prb,twocolumn,showpacs,amsmath,amssymb]{revtex4}

\usepackage{graphicx}
\usepackage{dcolumn}
\usepackage{bm}
\begin{document}

\title{Initial dissipation and current-voltage characteristics
of superconductors\\ containing fractal clusters of a normal
phase}

\author{Yuriy I. Kuzmin}
 \altaffiliation[Also at ]{Physics Department, State
Electrotechnical University, 5 Professor Popov Street, Saint
Petersburg 197376, Russia}
 \email{yurk@mail.ioffe.ru}

\affiliation{Ioffe Physical Technical Institute of the Russian
Academy of Sciences, 26 Polytechnicheskaya Street, Saint
Petersburg 194021 Russia}

\author{Mladen Prester}
 \email{prester@ifs.hr}
\affiliation{Institute of Physics, P.O. Box 304, HR-10000, Zagreb,
Croatia}

\date{\today}

\begin{abstract}
The influence of fractal clusters of a normal phase on distinctive
features of current-voltage (\textit{U}-\textit{I}) characteristic
of percolative type-II superconductors is considered. The results
of high-resolution measurements of the differential resistance of
BPSCCO/Ag composites are discussed in the context of magnetic flux
dynamics. The region of initial dissipation is observed on
\textit{U}-\textit{I} characteristics in the neighborhood of the
transition into a resistive state. In the course of this stage of
resistive transition the vortices start to break away from the
normal-phase clusters, which act as pinning centers. The effect of
transport current on vortex depinning is investigated.
A broad current range of initial dissipation
is considered as an evidence of fractal nature of the normal-phase clusters.
\end{abstract}

\pacs{74.81.-g; 74.25.Fy; 74.81.Bd}

\maketitle

\section{Introduction}

Superconductors containing fractal clusters of a normal phase have
specific magnetic and transport properties.
\cite{PresterPRB2,KuzminPLA1,KuzminPRB} The study of their
\textit{U}-\textit{I} characteristics enable to get new
information on the electromagnetic properties as well as on the
nature of a vortex state in such materials.
\cite{Blatter,Fisher,Fisher2,Brown,Yamafuji} The neighborhood of
resistive transition, especially its initial stage where the
energy dissipation sets-in, is of special interest. In this region
the process of vortex depinning gradually accrues resulting
finally with the destruction of a superconducting state.

The problem of initial dissipation in high-temperature
superconductors (HTS's) has been studied by many authors.
\cite{PresterPRB2,Fukumoto,Suenaga,Fukumoto2,Hlasnik,Polak,PresterPRB1,PresterSST,PresterSPIE}
The residual resistance of bismuth-based HTS's
[Bi$_{2}$Sr$_{2}$Ca$_{2}$Cu$_{3}$O$_{10+y}$ (BSCCO-2223) and
Bi$_{2}$Sr$_{2}$CaCu$_{2}$O$_{10+y}$ (BSCCO-2212)] at small
currents has been explained by deterioration of grain boundaries,
by initiation of micro-cracks which can act as chains of weak
links, \cite{Fukumoto,Suenaga} as well as by the grain-to-grain
misalignment or by degradation of the grains themselves
\cite{Fukumoto2}. A strong influence
of transfer resistance between superconducting and normal metal on \textit{U}%
-\textit{I} characteristics of silver-sheathed BSCCO-2223 tapes
has been found. \cite{Hlasnik} The ohmic behavior of
\textit{U}-\textit{I} curve on the initial stage of resistive
transition in BSCCO-2223 and BSCCO-2212 has been attributed to the
local transfer of excess current into the normal metal inclusions.
\cite{Polak} The fractal regime in the initial stage of
dissipation has been observed in BSCCO-2223,
YBa$_{2}$Cu$_{3}$O$_{7-x}$, and GdBa$_{2}$Cu$_{3}$O$_{7-x}$.
\cite{PresterPRB2} The fractal nature of the normal-phase clusters
contained in YBa$_{2}$Cu$_{3}$O$_{7-x}$ thin films has been found
\cite{KuzminPLA1} and the effect of such clusters on vortex
dynamics has been analyzed.
\cite{KuzminPRB,KuzminPLA2,KuzminPLA3,KuzminJLTP} In the present
work we consider the influence of the fractal clusters of a normal
phase on the initial part of \textit{U}-\textit{I} characteristic
near the resistive transition.

\section{Statement of a Problem}

Let us consider a superconductor containing inclusions of a normal
phase, which are out of contact with one another. We will suppose
that the characteristic sizes of these inclusions far exceed both
the superconducting coherence length and the penetration depth. A
prototype of such a structure is the superconducting wire or tape
armored by normal metal for giving the necessary electrical and
mechanical properties. \cite{Polak,Pashitski,Kovac} Concrete
example is the silver-sheathed HTS bismuth-based composites, which
are of practical interest for energy transport and storage.
\cite{Fukumoto,Hlasnik,PresterSPIE}

When electric current is passed through such a material, it flows
through a superconducting percolative cluster. A specificity of
the problem is that the percolative cluster consists of mesoscopic
superconducting islands joined by weak links.
\cite{PresterPRB2,PresterPRB1} As the transport current is
increased, the local currents flowing through ones or other weak
links begin to exceed the critical values, so some part of them
become resistive. Thus, the number of weak links involved in the
superconducting cluster is randomly reduced so the transition of a
superconductor into a resistive state corresponds to breaking of
the percolation through a superconducting cluster. The transport
current acts as a random generator that changes the relative
fractions of conducting components in classical percolative
medium, \cite{Stauffer} hence the resistive transition can be
treated as a current-induced critical phenomenon.
\cite{PresterPRB1,PresterSST}

On the other hand, the dissipation of energy in a superconductor
is linked with the vortex dynamics as any motion of a magnetic
flux induces an electric field. In HTS's the vortex motion is of
special importance because of large thermal fluctuations existing
at high temperatures and small pinning energies. \cite{Blatter}
The magnetic flux can move only after the vortices will be broken
away from the pinning centers. Until the moment when the Lorentz
force created by a transport current will exceed the pinning force
the magnetic flux remains trapped in normal-phase clusters. These
clusters present the sets of normal-phase inclusions, united be
the common trapped flux and surrounded by the superconducting
phase. \cite{KuzminPLA1,KuzminPRB} In such a system depinning has
a percolative character, \cite{Yamafuji,Ziese,Ziese2} because
vortices move through the randomly generated channels, connecting
the normal-phase clusters. Such channels of vortex transport can
be created by weak links, \cite{Duran} which form readily in HTS's
due to the intrinsically short coherence length. \cite{Sonier}
Depending on the specific weak link configuration each
normal-phase cluster has its own current of depinning, which
contributes to the total statistical distribution of critical
currents. Thus, the weak links do not only connect superconducting
clusters between themselves, maintaining the electrical current
percolation, but they also form the channels for vortex transport
so providing for the percolation of magnetic flux.

The depinning current of each cluster is related to the cluster
size, because the larger cluster has more weak links over its
boundary with the surrounding superconducting space, and thus the
smaller current of depinning. \cite{KuzminPRB} As a measure of the
cluster size we will take the area of its cross-section, and in
the subsequent text we will call this value simply ``the cluster
area``. In the general case the distribution of the cluster areas
may be described by gamma distribution with the probability
density
\begin{equation}
w\left( a\right) =\frac{\left( g+1\right) ^{g+1}}{\Gamma \left( g+1\right) }%
a^{g}\exp \left[ -\left( g+1\right) a\right]   \label{gamma1}
\end{equation}
where $\Gamma \left( \nu \right) $ is Euler gamma function, $a\equiv A/%
\overline{A}$\ is the dimensionless area of the cluster, $A$ is
the area of the cross-section of the cluster by the plane,
transversally to which the vortices are moving, $A_{0}>0$ and
$g>-1$ are the parameters of gamma
distribution that control the mean area of the cluster $\overline{A}%
=(g+1)A_{0}$ and its variance $\sigma _{A}^{2}=\left( g+1\right)
A_{0}^{2}$. The mean dimensionless area of the cluster is equal to
unity, whereas the variance is determined by $g$-parameter only:
$\sigma _{a}^{2}=1/\left( g+1\right) $.

In order to use the formulas for \textit{U}-\textit{I}
characteristics obtained in
Refs.~\onlinecite{KuzminPLA2,KuzminJLTP} for the case of thin
films with clusters of columnar defects, it is necessary to assume
that all the entry points into weak links are distributed
uniformly along extended parts of the normal-phase inclusions. The
problem setting for armored superconducting wires has some
distinctive features: (i) the fragments of a normal phase has the
form of extended inclusions oriented along an axis of the wire;
(ii) the magnetic flux is created by the transport current itself
and is concentrated along irregular-shaped rings, which are
deformed in such a way that the normal-phase clusters would be
most captured; (iii) the vortices are transferred through the weak
links connecting long parts of the normal-phase inclusions between
themselves.

Gamma distribution of the cluster area of Eq.~(\ref{gamma1}) gives
rise to exponential-hyperbolic distribution of depinning currents,
\cite{KuzminPLA3}
for which \textit{U}-\textit{I} characteristic has the form \cite{KuzminJLTP}%
:

\begin{eqnarray}
u &=&\frac{r_{f}}{\Gamma \left( g+1\right) }{\Biggl[}i\Gamma
\left(
g+1,G\,i^{-2/D}\right)  \nonumber \\
&&-G^{D/2}\Gamma \left( g+1-\frac{D}{2},G\,i^{-2/D}\right)
{\Biggr]} \label{volt2}
\end{eqnarray}
where
\[
G\equiv \left[ \frac{\theta ^{\theta }}{\theta ^{g+1}-\left(
D/2\right) \exp \left( \theta \right) \Gamma (g+1,\theta )}\right]
^{\frac{2}{D}}
\]
$\theta \equiv g+1+D/2$, $i\equiv I/I_{c}$ is the dimensionless
electrical current normalized to the critical current of the
transition into a resistive state $I_{c}=\alpha \left(
A_{0}G\right) ^{-D/2}$, $\alpha $ is the form factor, $D$ is the
fractal dimension of the cluster boundary, $\Gamma \left( \nu
,z\right) $ is the complementary incomplete gamma function. The
voltage across a sample $U$ and flux flow resistance $R_{f}$ are
linked to the corresponding dimensionless quantities $u$ and
$r_{f}$ by the formula: $U/R_{f}=I_{c}\left( u/r_{f}\right) $.

\begin{figure}
\includegraphics{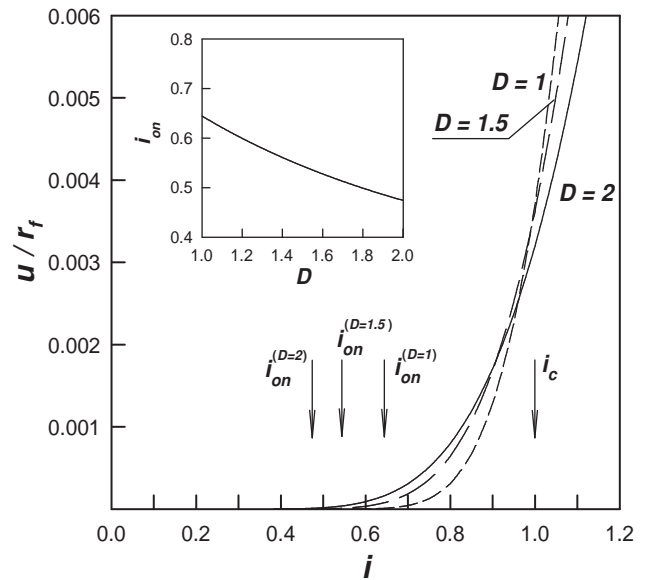}
\caption{\label{fig1} Initial region of \textit{U}-\textit{I}
characteristics of superconductor containing fractal clusters of a
normal phase. The arrows
show the critical current $i_{c}$ and the values of the onset current $%
i_{on} $ at the voltage resolution of $10^{-5}u/r_{f}$. The
dependence of the onset current on the fractal dimension of the
cluster boundary is presented in the inset.}
\end{figure}

The \textit{U}-\textit{I} characteristics in the simplest case of
$g=0$, when gamma distribution of Eq.~(\ref{gamma1}) is reduced to
the exponential one, $w\left( a\right) =\exp \left( -a\right) $,
are shown in Fig.~\ref{fig1}.
This figure demonstrates that in the range of the currents $i>1$
the fractality of the clusters reduces the voltage arising from
magnetic flux motion. Meanwhile, the situation is different below
the critical current. When $i<1$, the higher the fractal dimension
of the normal phase cluster is, the larger is the voltage across a
sample and the more stretched is the region of initial dissipation
on \textit{U}-\textit{I} characteristic. For further consideration
it is convenient to introduce the onset current $i_{on}$, from
which this region spreads away. The magnitude of this current is
set by the resolution of
voltage measurement. In Fig.~\ref{fig1} the arrows show the onset current values $%
i_{on}$ corresponding to the resolution level of $10^{-5}u/r_{f}$.
The inset of this figure shows the dependence of the onset current
$i_{on}$ on the fractal dimension.

\section{Experiment}

For experimental study of initial dissipation the high-resolution
measurements of differential resistance of HTS composites (BiPb)$_{2}$Sr$%
_{2} $Ca$_{2}$Cu$_{3}$O$_{10+y}$ containing inclusions of silver
(BPSCCO/Ag) were carried out. The samples had the form of
silver-sheathed tapes with superconducting core. These tapes were
prepared following a conventional ``powder-in-tube`` technique
\cite{Kovac2}. Magneto-optical micrographs of such tapes are given in
Refs.~\onlinecite{Polak,PresterSST,Pashitski}. The superconducting
core inevitably contains the normal-phase inclusions of Ag as well
as different secondary phases (the degraded non-stoichiometric
material, grain boundaries, voids, cracks, impurities, etc.). The
volume content of the normal phase in the core was far below the
percolation threshold so the percolative superconducting cluster
was dense enough. Comprehensive characterization of structural
(XRD, EDX, SEM, and TEM data), mechanical (microhardness) and
electrical properties (transport critical current) of the samples
can be found in Ref.~\onlinecite{Kovac}.

\textit{U}-\textit{I} characteristics were obtained by integration
of experimental dependencies of the differential resistance on a
current. Such a measurement technique was chosen for the following
reasons: (i) the differential resistance presents a small-signal
parameter suitable for description of the nonlinear
\textit{U}-\textit{I} characteristic; (ii) the lock-in \textit{ac}
technique provides the high resolution necessary to observe
peculiarities of the resistive transition. In our experiments the
differential resistance resolution was equal to 1\thinspace $\mu
\Omega $ that corresponds to the equivalent voltage resolution of
1.5\thinspace nV at \textit{ac} current component of 1.5\thinspace
mA. The measured differential resistance $R_{d}$ is proportional
to density of
vortices, $n$, broken away from the pinning centers: $%
R_{d}=nR_{f}\Phi _{0}/B$, where $B$ is the magnetic field, $\Phi
_{0}\equiv
hc/\left( 2e\right) $ is the magnetic flux quantum, $h$ is Planck constant, $%
c$ is the velocity of light, and $e$ is the electron charge. It is
just a motion of these vortices induces electrical field in a
sample and thus causes the dissipation.

\begin{figure}
\includegraphics{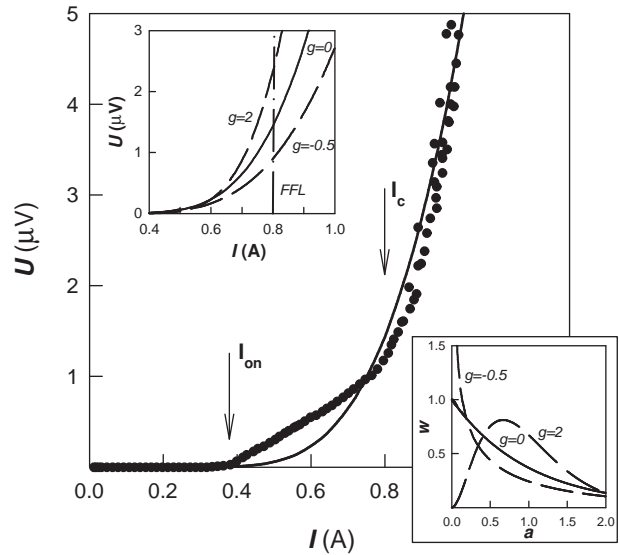}
\caption{\label{fig2} Experimental \textit{U}-\textit{I}
characteristic of HTS composite
BPSCCO/Ag (points) and theoretical curve calculated for fractal dimension $%
D=2$ in the case of exponential distribution of cluster areas
($g=0$). The arrows show the onset current $I_{on}$ and the
critical current of resistive transition $I_{c}$. In the lower
inset the probability density for gamma distribution of
normal-phase cluster areas is presented, and in the upper inset
the corresponding \textit{U}-\textit{I} characteristics for
different values of $g$-parametes are given. Dash-dot line in the
upper insert, almost vertical due to enlarged voltage scale,
corresponds to the case of flux flow limit (\textit{FFL}).}
\end{figure}

The obvious region of initial dissipation was observed in ten
samples taken from different batches. The \textit{U}-\textit{I}
characteristic of one of the samples, where this region was more
clearly defined, is presented in Fig.~\ref{fig2}.
The parameters of this sample were as follows: cross-section area
of the superconducting core was equal to 0.068\thinspace mm$^{2}$
at the thickness of 30\thinspace $\mu $m while the width of the
tape was 4\thinspace mm at the thickness of 0.2\thinspace mm, transition temperature was $%
T_{c}$=107.3\thinspace K, normal state resistance just above the transition (at $T$=110%
\thinspace K) was equal to 545\thinspace $\mu \Omega $, critical
current density at $T$=77\thinspace K (criterion 1\thinspace $\mu
V/cm$, self-field) was equal to 3.1\thinspace kA/cm$^{2}$. The
measurements were carried out at temperature of 96.5\thinspace K
in the Earth magnetic field. Notice that this temperature was more
than 10\thinspace K below he transition temperature so any marked
self-heating effects are excluded. The tape form of our samples
provides an ideal thermal contact with the thermally stabilized
copper substrate so the Joule heating at the current contacts was
substantially diminished. The measurements were also performed in
the temperature range 90\thinspace K$<T<$105\thinspace K, and in
applied magnetic field up to 100\thinspace Oe, but obtained
results were similar. The flux flow resistance, which gives the
asymptotic slope of \textit{U}-\textit{I} characteristic of a
superconductor in a resistive state, was equal to
$R_{f}=560\thinspace \,\mu \Omega $. The corresponding curve in
the limiting case of flux flow is shown by the dash-dot line in
the upper inset of Fig.~\ref{fig2}. The value of the critical
current of resistive transition $I_{c}$=0.8\thinspace A was found
by the point of intersection of abscissa axis and the tangent line
drawn through the inflection point of the curve of dependence of
the differential resistance on the current. Theoretical
\textit{U}-\textit{I} characteristic, found from
equation~(\ref{volt2}), is shown in Fig.~\ref{fig2} by a solid
line. In order to highlight the effects associated with the
fractal properties of clusters, the greatest possible value of
fractal dimension $D=2$ was taken (such a
fractal dimension is inherent, for example, in Peano curves. \cite{Mandelbrot}%
) In this case the onset current, calculated on the level of
$10^{-5}u/r_{f}$, is equal to $I_{on}$=0.38\thinspace A (see also
inset of
Fig.~\ref{fig1}). The only one free parameter of the theoretical model is $g$%
-parameter of gamma distribution. Inasmuch as there is no
experimental information about statistical distribution of
normal-phase cluster areas, we have taken the simplest case of
exponential distribution ($g=0$). The insets of Fig.~\ref{fig2}
show how the distribution of the cluster sizes (lower inset) can
affect the \textit{U}-\textit{I} characteristics (upper inset).
The theoretical curve agrees with experimental data in the main
point: the \textit{U}-\textit{I} characteristic does not start
exactly from the critical current $I_{c}$, but there is a
well-defined initial region that begins with the onset current
$I_{on}$, with the perfect coincidence between the calculated and
measured magnitudes of latter parameter. At the same time, the
theoretical curve passes below the experimental points at small
currents and has no bend near $I_{c}$. This feature may be
explained by that the theoretical curve was calculated for the
case of the constant value of fractal dimension, whereas the
experimental data relate to the situation where the fractal
dimension is not the same for the different parts of
\textit{U}-\textit{I} characteristic because its value is governed
by the transport current. Indeed, on the current-induced resistive
transition the cluster topology varies as the transport current
increases. As more and more weak links turn resistive, the
superconducting cluster becomes less and less ramified. This can
result in changing in fractal dimension of the normal-phase
clusters as the boundary between superconducting and normal phases
becomes less indented. Thus it is necessary to take into account
that the interface sweeps through the area of reduced value of a
superconducting order parameter that surrounds the normal-phase
inclusions, so its position depends on the magnitude of transport
current. Therefore dependence of the fractal dimension on the
current can be very complicated. In addition to that, the
differences observed may be ascribed to the peculiarities of flux
creep in the initial stage of resistive transition, which are not
taken into account within the used theoretical model.

In conclusion, we have shown in this paper that the initial region
of dissipation in HTS's can be considered as vortex dynamics
phenomenon, which demonstrates the fractal properties of the
normal-phase clusters.

\begin{acknowledgments}
The work is carried out at support of Russian Foundation for Basic
Researches (grant No 02-02-17667) and Croatian Ministry of Science
and Technology (Project No 0035009). M.P. also acknowledges
support of the Swiss National Science Foundation, SCOPES project
No 7KRPJ065631.
\end{acknowledgments}

\bibliography{Fractal}

\begin{thebibliography}{28}
\expandafter\ifx\csname natexlab\endcsname\relax\def\natexlab#1{#1}\fi
\expandafter\ifx\csname bibnamefont\endcsname\relax
  \def\bibnamefont#1{#1}\fi
\expandafter\ifx\csname bibfnamefont\endcsname\relax
  \def\bibfnamefont#1{#1}\fi
\expandafter\ifx\csname citenamefont\endcsname\relax
  \def\citenamefont#1{#1}\fi
\expandafter\ifx\csname url\endcsname\relax
  \def\url#1{\texttt{#1}}\fi
\expandafter\ifx\csname urlprefix\endcsname\relax\def\urlprefix{URL }\fi
\providecommand{\bibinfo}[2]{#2}
\providecommand{\eprint}[2][]{\url{#2}}

\bibitem[{\citenamefont{Prester}(1999)}]{PresterPRB2}
\bibinfo{author}{\bibfnamefont{M.}~\bibnamefont{Prester}},
  \bibinfo{journal}{Phys.\ Rev.\ B} \textbf{\bibinfo{volume}{60}},
  \bibinfo{pages}{3100} (\bibinfo{year}{1999}).

\bibitem[{\citenamefont{Kuzmin}(2000)}]{KuzminPLA1}
\bibinfo{author}{\bibfnamefont{Y.~I.} \bibnamefont{Kuzmin}},
  \bibinfo{journal}{Phys.\ Lett.\ A} \textbf{\bibinfo{volume}{267}},
  \bibinfo{pages}{66} (\bibinfo{year}{2000}).

\bibitem[{\citenamefont{Kuzmin}(2001{\natexlab{a}})}]{KuzminPRB}
\bibinfo{author}{\bibfnamefont{Y.~I.} \bibnamefont{Kuzmin}},
  \bibinfo{journal}{Phys.\ Rev.\ B} \textbf{\bibinfo{volume}{64}},
  \bibinfo{pages}{094519} (\bibinfo{year}{2001}{\natexlab{a}}).

\bibitem[{\citenamefont{Blatter et~al.}(1994)\citenamefont{Blatter, Feigelman,
  Geshkenbein, Larkin, and Vinokur}}]{Blatter}
\bibinfo{author}{\bibfnamefont{G.}~\bibnamefont{Blatter}},
  \bibinfo{author}{\bibfnamefont{M.~V.} \bibnamefont{Feigelman}},
  \bibinfo{author}{\bibfnamefont{V.~B.} \bibnamefont{Geshkenbein}},
  \bibinfo{author}{\bibfnamefont{A.~I.} \bibnamefont{Larkin}},
  \bibnamefont{and} \bibinfo{author}{\bibfnamefont{V.~M.}
  \bibnamefont{Vinokur}}, \bibinfo{journal}{Rev.\ Mod.\ Phys.}
  \textbf{\bibinfo{volume}{66}}, \bibinfo{pages}{1125} (\bibinfo{year}{1994}).

\bibitem[{\citenamefont{Fisher}(1989)}]{Fisher}
\bibinfo{author}{\bibfnamefont{M.~P.~A.} \bibnamefont{Fisher}},
  \bibinfo{journal}{Phys.\ Rev.\ Lett.} \textbf{\bibinfo{volume}{62}},
  \bibinfo{pages}{1415} (\bibinfo{year}{1989}).

\bibitem[{\citenamefont{Fisher et~al.}(1991)\citenamefont{Fisher, Fisher, and
  Huse}}]{Fisher2}
\bibinfo{author}{\bibfnamefont{D.~S.} \bibnamefont{Fisher}},
  \bibinfo{author}{\bibfnamefont{M.~P.~A.} \bibnamefont{Fisher}},
  \bibnamefont{and} \bibinfo{author}{\bibfnamefont{D.~A.} \bibnamefont{Huse}},
  \bibinfo{journal}{Phys.\ Rev.\ B} \textbf{\bibinfo{volume}{43}},
  \bibinfo{pages}{130} (\bibinfo{year}{1991}).

\bibitem[{\citenamefont{Brown}(2000)}]{Brown}
\bibinfo{author}{\bibfnamefont{B.}~\bibnamefont{Brown}},
  \bibinfo{journal}{Phys.\ Rev.\ B} \textbf{\bibinfo{volume}{61}},
  \bibinfo{pages}{3267} (\bibinfo{year}{2000}).

\bibitem[{\citenamefont{Yamafuji and Kiss}(1997)}]{Yamafuji}
\bibinfo{author}{\bibfnamefont{K.}~\bibnamefont{Yamafuji}} \bibnamefont{and}
  \bibinfo{author}{\bibfnamefont{T.}~\bibnamefont{Kiss}},
  \bibinfo{journal}{Physica C} \textbf{\bibinfo{volume}{290}},
  \bibinfo{pages}{9} (\bibinfo{year}{1997}).

\bibitem[{\citenamefont{Fukumoto et~al.}(1995)\citenamefont{Fukumoto, Li, Wang,
  Suenaga, and Haldar}}]{Fukumoto}
\bibinfo{author}{\bibfnamefont{Y.}~\bibnamefont{Fukumoto}},
  \bibinfo{author}{\bibfnamefont{Q.}~\bibnamefont{Li}},
  \bibinfo{author}{\bibfnamefont{Y.~L.} \bibnamefont{Wang}},
  \bibinfo{author}{\bibfnamefont{M.}~\bibnamefont{Suenaga}}, \bibnamefont{and}
  \bibinfo{author}{\bibfnamefont{P.}~\bibnamefont{Haldar}},
  \bibinfo{journal}{Appl.\ Phys.\ Lett.} \textbf{\bibinfo{volume}{66}},
  \bibinfo{pages}{1827} (\bibinfo{year}{1995}).

\bibitem[{\citenamefont{Suenaga et~al.}(1995)\citenamefont{Suenaga, Fukumoto,
  Haldar, Thurston, and Wildgruber}}]{Suenaga}
\bibinfo{author}{\bibfnamefont{M.}~\bibnamefont{Suenaga}},
  \bibinfo{author}{\bibfnamefont{Y.}~\bibnamefont{Fukumoto}},
  \bibinfo{author}{\bibfnamefont{P.}~\bibnamefont{Haldar}},
  \bibinfo{author}{\bibfnamefont{T.~R.} \bibnamefont{Thurston}},
  \bibnamefont{and}
  \bibinfo{author}{\bibfnamefont{U.}~\bibnamefont{Wildgruber}},
  \bibinfo{journal}{Appl.\ Phys.\ Lett.} \textbf{\bibinfo{volume}{67}},
  \bibinfo{pages}{3025} (\bibinfo{year}{1995}).

\bibitem[{\citenamefont{Fukumoto et~al.}(1996)\citenamefont{Fukumoto,
  Moodenbaugh, Suenaga, Fischer, Shibutani, Hase, and Hayashi}}]{Fukumoto2}
\bibinfo{author}{\bibfnamefont{Y.}~\bibnamefont{Fukumoto}},
  \bibinfo{author}{\bibfnamefont{A.~R.} \bibnamefont{Moodenbaugh}},
  \bibinfo{author}{\bibfnamefont{M.}~\bibnamefont{Suenaga}},
  \bibinfo{author}{\bibfnamefont{D.~A.} \bibnamefont{Fischer}},
  \bibinfo{author}{\bibfnamefont{K.}~\bibnamefont{Shibutani}},
  \bibinfo{author}{\bibfnamefont{T.}~\bibnamefont{Hase}}, \bibnamefont{and}
  \bibinfo{author}{\bibfnamefont{S.}~\bibnamefont{Hayashi}},
  \bibinfo{journal}{J.\ Appl.\ Phys.} \textbf{\bibinfo{volume}{80}},
  \bibinfo{pages}{331} (\bibinfo{year}{1996}).

\bibitem[{\citenamefont{Hlasnik et~al.}(1996)\citenamefont{Hlasnik, Jansak,
  Majoros, Kokavec, and Chovanec}}]{Hlasnik}
\bibinfo{author}{\bibfnamefont{I.}~\bibnamefont{Hlasnik}},
  \bibinfo{author}{\bibfnamefont{L.}~\bibnamefont{Jansak}},
  \bibinfo{author}{\bibfnamefont{M.}~\bibnamefont{Majoros}},
  \bibinfo{author}{\bibfnamefont{J.}~\bibnamefont{Kokavec}}, \bibnamefont{and}
  \bibinfo{author}{\bibfnamefont{F.}~\bibnamefont{Chovanec}},
  \bibinfo{journal}{IEEE\ Trans.\ Magnetics} \textbf{\bibinfo{volume}{32}},
  \bibinfo{pages}{2806} (\bibinfo{year}{1996}).

\bibitem[{\citenamefont{Polak et~al.}(1997)\citenamefont{Polak, Zhang, Parrell,
  Cai, Polyanskii, Hellstrom, Larbalestier, and Majoros}}]{Polak}
\bibinfo{author}{\bibfnamefont{M.}~\bibnamefont{Polak}},
  \bibinfo{author}{\bibfnamefont{W.}~\bibnamefont{Zhang}},
  \bibinfo{author}{\bibfnamefont{J.}~\bibnamefont{Parrell}},
  \bibinfo{author}{\bibfnamefont{X.~Y.} \bibnamefont{Cai}},
  \bibinfo{author}{\bibfnamefont{A.}~\bibnamefont{Polyanskii}},
  \bibinfo{author}{\bibfnamefont{E.~E.} \bibnamefont{Hellstrom}},
  \bibinfo{author}{\bibfnamefont{D.~C.} \bibnamefont{Larbalestier}},
  \bibnamefont{and} \bibinfo{author}{\bibfnamefont{M.}~\bibnamefont{Majoros}},
  \bibinfo{journal}{Supercond.\ Sci.\ Technol.} \textbf{\bibinfo{volume}{10}},
  \bibinfo{pages}{769} (\bibinfo{year}{1997}).

\bibitem[{\citenamefont{Prester}(1996)}]{PresterPRB1}
\bibinfo{author}{\bibfnamefont{M.}~\bibnamefont{Prester}},
  \bibinfo{journal}{Phys.\ Rev.\ B} \textbf{\bibinfo{volume}{54}},
  \bibinfo{pages}{606} (\bibinfo{year}{1996}).

\bibitem[{\citenamefont{Prester}(1998)}]{PresterSST}
\bibinfo{author}{\bibfnamefont{M.}~\bibnamefont{Prester}},
  \bibinfo{journal}{Supercond.\ Sci.\ Technol.} \textbf{\bibinfo{volume}{11}},
  \bibinfo{pages}{333} (\bibinfo{year}{1998}).

\bibitem[{\citenamefont{Prester et~al.}(1998)\citenamefont{Prester,
  Kov\'{a}\v{c}, and Hu\v{s}ek}}]{PresterSPIE}
\bibinfo{author}{\bibfnamefont{M.}~\bibnamefont{Prester}},
  \bibinfo{author}{\bibfnamefont{P.}~\bibnamefont{Kov\'{a}\v{c}}},
  \bibnamefont{and}
  \bibinfo{author}{\bibfnamefont{I.}~\bibnamefont{Hu\v{s}ek}},
  \bibinfo{journal}{Proc.\ SPIE} \textbf{\bibinfo{volume}{3481}},
  \bibinfo{pages}{60} (\bibinfo{year}{1998}).

\bibitem[{\citenamefont{Kuzmin}(2001{\natexlab{b}})}]{KuzminPLA2}
\bibinfo{author}{\bibfnamefont{Y.~I.} \bibnamefont{Kuzmin}},
  \bibinfo{journal}{Phys.\ Lett.\ A} \textbf{\bibinfo{volume}{281}},
  \bibinfo{pages}{39} (\bibinfo{year}{2001}{\natexlab{b}}).

\bibitem[{\citenamefont{Kuzmin}(2002)}]{KuzminPLA3}
\bibinfo{author}{\bibfnamefont{Y.~I.} \bibnamefont{Kuzmin}},
  \bibinfo{journal}{Phys.\ Lett.\ A} \textbf{\bibinfo{volume}{300}},
  \bibinfo{pages}{510} (\bibinfo{year}{2002}).

\bibitem[{\citenamefont{Kuzmin}(2003)}]{KuzminJLTP}
\bibinfo{author}{\bibfnamefont{Y.~I.} \bibnamefont{Kuzmin}},
  \bibinfo{journal}{J.\ Low\ Temp.\ Phys.} \textbf{\bibinfo{volume}{130}},
  \bibinfo{pages}{261} (\bibinfo{year}{2003}).

\bibitem[{\citenamefont{Pashitski et~al.}(1995)\citenamefont{Pashitski,
  Polyanskii, Gurevich, Parrell, and Larbalestier}}]{Pashitski}
\bibinfo{author}{\bibfnamefont{A.~E.} \bibnamefont{Pashitski}},
  \bibinfo{author}{\bibfnamefont{A.}~\bibnamefont{Polyanskii}},
  \bibinfo{author}{\bibfnamefont{A.}~\bibnamefont{Gurevich}},
  \bibinfo{author}{\bibfnamefont{J.~A.} \bibnamefont{Parrell}},
  \bibnamefont{and} \bibinfo{author}{\bibfnamefont{D.~C.}
  \bibnamefont{Larbalestier}}, \bibinfo{journal}{Physica C}
  \textbf{\bibinfo{volume}{246}}, \bibinfo{pages}{133} (\bibinfo{year}{1995}).

\bibitem[{\citenamefont{Kov\'{a}\v{c} et~al.}(1997)\citenamefont{Kov\'{a}\v{c},
  Eastell, Pachla, Hu\v{s}ek, Marciniak, Grovenor, and Goringe}}]{Kovac}
\bibinfo{author}{\bibfnamefont{P.}~\bibnamefont{Kov\'{a}\v{c}}},
  \bibinfo{author}{\bibfnamefont{C.~J.} \bibnamefont{Eastell}},
  \bibinfo{author}{\bibfnamefont{W.}~\bibnamefont{Pachla}},
  \bibinfo{author}{\bibfnamefont{I.}~\bibnamefont{Hu\v{s}ek}},
  \bibinfo{author}{\bibfnamefont{H.}~\bibnamefont{Marciniak}},
  \bibinfo{author}{\bibfnamefont{C.~R.~M.} \bibnamefont{Grovenor}},
  \bibnamefont{and} \bibinfo{author}{\bibfnamefont{M.~J.}
  \bibnamefont{Goringe}}, \bibinfo{journal}{Physica C}
  \textbf{\bibinfo{volume}{292}}, \bibinfo{pages}{322} (\bibinfo{year}{1997}).

\bibitem[{\citenamefont{Stauffer}(1979)}]{Stauffer}
\bibinfo{author}{\bibfnamefont{D.}~\bibnamefont{Stauffer}},
  \bibinfo{journal}{Phys.\ Rep.} \textbf{\bibinfo{volume}{54}},
  \bibinfo{pages}{2} (\bibinfo{year}{1979}).

\bibitem[{\citenamefont{Ziese}(1996{\natexlab{a}})}]{Ziese}
\bibinfo{author}{\bibfnamefont{M.}~\bibnamefont{Ziese}},
  \bibinfo{journal}{Physica C} \textbf{\bibinfo{volume}{269}},
  \bibinfo{pages}{35} (\bibinfo{year}{1996}{\natexlab{a}}).

\bibitem[{\citenamefont{Ziese}(1996{\natexlab{b}})}]{Ziese2}
\bibinfo{author}{\bibfnamefont{M.}~\bibnamefont{Ziese}},
  \bibinfo{journal}{Phys.\ Rev.\ B} \textbf{\bibinfo{volume}{53}},
  \bibinfo{pages}{12422} (\bibinfo{year}{1996}{\natexlab{b}}).

\bibitem[{\citenamefont{Dur\'{a}n et~al.}(1992)\citenamefont{Dur\'{a}n, Gammel,
  Wolfe, Fratello, Bishop, Rice, and Ginsberg}}]{Duran}
\bibinfo{author}{\bibfnamefont{C.~A.} \bibnamefont{Dur\'{a}n}},
  \bibinfo{author}{\bibfnamefont{P.~L.} \bibnamefont{Gammel}},
  \bibinfo{author}{\bibfnamefont{R.}~\bibnamefont{Wolfe}},
  \bibinfo{author}{\bibfnamefont{V.~J.} \bibnamefont{Fratello}},
  \bibinfo{author}{\bibfnamefont{D.~J.} \bibnamefont{Bishop}},
  \bibinfo{author}{\bibfnamefont{J.~P.} \bibnamefont{Rice}}, \bibnamefont{and}
  \bibinfo{author}{\bibfnamefont{D.~M.} \bibnamefont{Ginsberg}},
  \bibinfo{journal}{Nature (London)} \textbf{\bibinfo{volume}{357}},
  \bibinfo{pages}{474} (\bibinfo{year}{1992}).

\bibitem[{\citenamefont{Sonier et~al.}(1999)\citenamefont{Sonier, Kiefl,
  Brewer, Bonn, Dunsiger, Hardy, Liang, Miller, Noakes, and Stronach}}]{Sonier}
\bibinfo{author}{\bibfnamefont{J.~E.} \bibnamefont{Sonier}},
  \bibinfo{author}{\bibfnamefont{R.~F.} \bibnamefont{Kiefl}},
  \bibinfo{author}{\bibfnamefont{J.~H.} \bibnamefont{Brewer}},
  \bibinfo{author}{\bibfnamefont{D.~A.} \bibnamefont{Bonn}},
  \bibinfo{author}{\bibfnamefont{S.~R.} \bibnamefont{Dunsiger}},
  \bibinfo{author}{\bibfnamefont{W.~N.} \bibnamefont{Hardy}},
  \bibinfo{author}{\bibfnamefont{R.}~\bibnamefont{Liang}},
  \bibinfo{author}{\bibfnamefont{R.~I.} \bibnamefont{Miller}},
  \bibinfo{author}{\bibfnamefont{D.~R.} \bibnamefont{Noakes}},
  \bibnamefont{and} \bibinfo{author}{\bibfnamefont{C.~E.}
  \bibnamefont{Stronach}}, \bibinfo{journal}{Phys.\ Rev.\ B}
  \textbf{\bibinfo{volume}{59}}, \bibinfo{pages}{R729} (\bibinfo{year}{1999}).

\bibitem[{\citenamefont{Kov\'{a}\v{c} et~al.}(1995)\citenamefont{Kov\'{a}\v{c},
  Hu\v{s}ek, Pachla, Meli\v{s}ek, and Kliment}}]{Kovac2}
\bibinfo{author}{\bibfnamefont{P.}~\bibnamefont{Kov\'{a}\v{c}}},
  \bibinfo{author}{\bibfnamefont{I.}~\bibnamefont{Hu\v{s}ek}},
  \bibinfo{author}{\bibfnamefont{W.}~\bibnamefont{Pachla}},
  \bibinfo{author}{\bibfnamefont{T.}~\bibnamefont{Meli\v{s}ek}},
  \bibnamefont{and} \bibinfo{author}{\bibfnamefont{V.}~\bibnamefont{Kliment}},
  \bibinfo{journal}{Supercond.\ Sci.\ Technol.} \textbf{\bibinfo{volume}{8}},
  \bibinfo{pages}{341} (\bibinfo{year}{1995}).

\bibitem[{\citenamefont{Mandelbrot}(1977)}]{Mandelbrot}
\bibinfo{author}{\bibfnamefont{B.~B.} \bibnamefont{Mandelbrot}},
  \emph{\bibinfo{title}{Fractals: Form, Chance, and Dimension}}
  (\bibinfo{publisher}{Freeman}, \bibinfo{address}{San Francisco},
  \bibinfo{year}{1977}).

\end{thebibliography}

\end{document}